\documentclass[prc,onecolumn,showpacs,floatfix,nofootinbib,preprintnumbers,superscriptaddress,amsmath,amssymb]{revtex4-2}

\usepackage{CJKutf8}
\usepackage{graphicx}
\usepackage{epsfig}
\usepackage{amsmath,amssymb}
\usepackage{bm}
\usepackage{color}
\usepackage{float}
\usepackage{dcolumn}
\usepackage{multirow}
\usepackage{mathrsfs}
\usepackage[arrowdel]{physics}
\usepackage[OT1]{fontenc}
\usepackage{outline}
\usepackage{tikz,tikz-3dplot}
\usetikzlibrary{positioning}

\DeclareMathAlphabet{\mathcal}{OMS}{cmsy}{m}{n}

\renewcommand{\vec}[1]{\mbox{\boldmath $#1$}}

\newcommand{\HFODD}{\textsc{hfodd}}

\begin{document}
\begin{CJK*}{UTF8}{gbsn}

\title{Solving the three-dimensional Skyrme Hartree-Fock-Bogoliubov problem using the mixed-basis method}

\author{Yue Shi (石跃)\thanks{corresponding author}}
\email[corresponding author: ]{yueshi@hit.edu.cn}
\affiliation{Department of Physics, Harbin Institute of Technology, Harbin 150001, People's Republic of China}

\author{Nobuo Hinohara}
\affiliation{Center for Computational Sciences, University of Tsukuba, Tsukuba 305-8577, Japan}
\affiliation{Faculty of Pure and Applied Sciences, University of Tsukuba, Tsukuba 305-8571, Japan}

\begin{abstract}

\begin{description}

\item[Background] The symmetry-unrestricted Hartree-Fock-Bogoliubov (HFB)
simulation is important for describing various quantum many-body systems.
However, the HFB problem in Cartesian coordinate space is numerically
challenging.

\item[Purpose] For describing ground states without imposing axial symmetry and
looking ahead to future extension for dynamics with full time dependence, we
present a numerically efficient implementation of the three-dimensional (3D)
HFB code.

\item[Methods] We develop a 3D Skyrme HFB code based on the mixed-basis
representation (HFB$^{\rm mix}$) which consists of two harmonic-oscillator (HO)
bases in the $x$- and $y$-directions, and finite-difference (FD) basis in the
$z$-direction in solving the nuclear 3D HFB problem.

\item[Results] The results show very well agreement among all the three codes
(HFB$^{\rm mix}$, HO3D, and \HFODD). Especially for the HF calculations, the
differences in total energies are on the order of a few keV for the lightest O
and Mg nuclei. The HFB$^{\rm mix}$ is applied to spherical, prolate, and
triaxial systems, and gives the same quadrupole moments for the deformed nuclei
as those of the HO-based calculations. Feasibility of the HFB$^{\rm mix}$ is
demonstrated in the fission isomer and barrier calculations of $^{240}$Pu.

\item[Conclusions] The HFB$^{\rm mix}$ is useful for solving the nuclear 3D HFB
problem for its numerical efficiency. Future work will include the analysis of
deformed drip-line systems and systematic potential-energy surface calculation
for fission-path analysis as well as the time-dependent extension of the
HFB$^{\rm mix}$ code for dynamics calculations.

\end{description}

\end{abstract}


\maketitle
\end{CJK*}

\section{Introduction}
\label{sec1}

For various quantum many-body simulations in quantum chemistry,
condensed-matter physics, atomic physics, and nuclear physics, one deals with
problems where constituent particles extend far spatially~\cite{naka16}. 
For example, inside a crystal lattice, the wave functions of the electrons spread extensively within 
the Coulomb potential exerted by the atomic nuclei which are distributed periodically.
Similarly, a composite molecule often constitutes many atoms situated at a significant distance. 
While the electrons of such a molecule may be localized around a certain atom in the ground state, 
they may be extended over the whole molecule in the excited states.  
For some drip-line nuclei where the Fermi energy is close to zero, the dilute nuclear material densities may extend
spatially when constituent nucleons occupy orbits with low orbital angular
momenta (halo nuclei). In these situations, a microscopic description for the
many-body wave functions of the quantum systems {\it requires} the solution in
a three-dimensional (3D) coordinate/momentum space.

In theoretical nuclear physics, a multitude of independent-particle (IP) models
play important roles. The conventional IP models use the well-tuned modified
harmonic oscillator (HO) or the Woods-Saxon potential plus the spin-orbit (SO)
potential to reproduce the experimental magic numbers~\cite{ring80}. Such
mean-field models together with the Strutinski shell correction
method~\cite{brack72} achieve quantitative success in describing a variety of
observables in low-energy nuclear physics. These models are known as
microscopic-macroscopic models.

Starting from an effective nucleon-nucleon interaction or Lagrangian together
with a variational Hartree-Fock (HF) procedure, one determines the one-body
potential in a self-consistent manner~\cite{bend03}. These IP models are known
as the self-consistent mean-field theories which are based on the zero-range
Skyrme~\cite{vaut72,doba84} or the finite-range
Gogny~\cite{gogny75,gogny80,gogny09} forces, and the relativistic mean-field
theory~\cite{serot92,ring96,meng06,shen19}. With the advances in synthesizing
the exotic nuclei with extreme $N/Z$ ratio, modern self-consistent mean-field
theories are more capable of accounting for new features which were not present
in nuclei near the valley of stability, such as halo nuclei mentioned above.
This {\it requires} the models to be flexible enough to work in a 3D coordinate
space, yet efficient enough for the beyond-mean-field extensions and to cover a
large parameter space~\cite{BONCHE1985,bend03}.

However, the requirements are, in general, numerically formidable with the
Hartree-Fock-Bogoliubov (HFB) treatment of the pairing correlation. Indeed, in
a typical nuclear 3D coordinate-space HFB simulation, one needs to repeatedly
diagonalize the Hamiltonian matrix of the dimension of a few hundreds of
thousands~\cite{pei14}. Recent application of the shifted Krylov method
\cite{jin17,kashiwaba19} and expansion in the three-dimensional wave-number
space basis \cite{yamagami2019} accelerate the 3D symmetry-unrestricted HFB
calculations. In comparison, one-~\cite{doba84,benn05,POSCHL1997} or
two-dimensional~\cite{teran03,pei08,yoshida08} HFB calculations in coordinate
space are less expensive due to their restricted spatial degrees of freedom.

In realistic applications, one frequently encounters situations where one- or
two-dimensional degrees of freedom play a major role in the static or dynamical
phenomena which do not possess any spatial symmetry. For example, a fissioning
nucleus assumes rather elongated deformation near or beyond the scission point.
The inner core of a neutron star may become so dense that the ``nuclei'' are
packed closed to each other with various exotic deformations favoring only one
or two spacial degrees of freedom. These are known as the nuclear
pasta~\cite{ravenhall1983,hashimoto1984} which have been discussed with various
Hartree-Fock calculations in coordinate space~\cite{magi02,newt09,schu15}. In
such situations it is more sensible to adopt a coordinate basis in the
elongated direction(s) and use HO bases for the remaining directions where the
matter is more restricted.

Indeed, a similar idea of using a mixed basis has been realized in the nuclear
finite-range Gogny HFB method and its time-dependent
extension~\cite{hash12,hash13}. The current work intends to describe the
implementation of the nuclear Skyrme HFB method using the HO plus
finite-difference (HO+FD) mixed basis. Throughout this work, we refer to the
basis which consists of the two HO bases in the $x$- and $y$-directions, and
the FD treatment in the $z$-direction as the ``mixed basis.''

In Sec.~\ref{model}, we describe the nuclear HFB problem and the numerical
details for the HO+FD mixed-basis HFB calculation. Section~\ref{results}
presents a systematic comparison for the results using the mixed-basis code and
those using a 3D HO based code (HO3D) developed in this work and the
\HFODD~codes (version 2.49t)~\cite{doba09}. The summary and perspective are
included in Sec.~\ref{summary}.

\section{The model}
\label{model}

In this section, we provide a short description of the nuclear HFB theory. We
then provide the numerical details for the implementations of the HF(B)$^{\rm
mix}$ code and the parameters chosen for the calculations.

\subsection{The Skyrme HFB formalism}
\label{HFB}

In the nuclear Skyrme HF mean-field theory, the total energy $E_{\rm Total}$ of a nucleus
can be decomposed into the kinetic, Skyrme, pairing, and Coulomb terms:
\begin{eqnarray}
\label{Etotal}
E_{\rm Total} &=& E_{\rm kin+c.m.} + E_{\rm Skyrme} + E_{\rm pair} + E_{\rm Coul} \nonumber \\
  &=& \int d^3r \big[\mathcal{K}(\vec{r}) + \mathcal{E}_{\rm Skyrme}(\vec{r}) \nonumber \\
  & & \quad + \mathcal{E}_{\rm pair}(\vec{r}) + \mathcal{E}_{\rm Coul}(\vec{r})\big].
\end{eqnarray}
The energy densities, $\mathcal{K}(\vec{r})$, $\mathcal{E}_{\rm Skyrme}(\vec{r})$, 
$\mathcal{E}_{\rm pair}(\vec{r})$, and $\mathcal{E}_{\rm Coul}(\vec{r})$, 
are functionals of various local densities (scalar density $\rho$, kinetic density $\tau$, 
spin-current density $\vec{\mathrm{J}}$, and pair density $\tilde{\rho}$). 
The kinetic-energy density of the nucleus with the one-body center-of-mass correction is 
given by
\begin{eqnarray}
\label{EDFkinetic}
\mathcal{K}(\bm{r}) = \frac{\hbar^2}{2m} \tau \qty(1-\frac{1}{A}),
\end{eqnarray}
where $m$ is the nucleon mass and $A$ is the mass number. 
The time-even part of the Skyrme energy density functional reads
\begin{eqnarray}
\label{skyrme}
\mathcal{E}_{\rm Skyrme}(\bm{r}) &=& \frac{b_0}{2} \rho^2 - \frac{b'_0}{2}\sum_q \rho^2_q + b_1\rho\tau - b'_1\sum_q\rho_q\tau_q \nonumber \\
              && \quad - \frac{b_2}{2} \rho\nabla^2\rho \nonumber + \frac{b'_2}{2}\sum_q\rho_q\nabla^2\rho_q \nonumber + \frac{b_3}{3}\rho^{\alpha+2} - \frac{b'_3}{3}\rho^{\alpha}\sum_q \rho^2_q \nonumber \\
              && \quad - b_4\rho\vec{\nabla}\cdot\vec{\mathrm{J}} - b'_4\sum_q\rho_q\vec{\nabla}\cdot\vec{\mathrm{J}}_q.
\end{eqnarray}
The index $q$ denotes neutrons ($n$) or protons ($p$). The local densities
without the index $q$ indicate the sum of neutron and proton densities. The
current work assumes the time-reversal symmetry and discusses only even-even
systems. Therefore, the time-odd part of the energy density functional does
not contribute in Eq.~(\ref{skyrme}).
The Coulomb energy is decomposed into the direct and exchange terms
\begin{equation}
\mathcal{E}_{\rm Coul}(\vec{r}) = \mathcal{E}_{\rm Coul}^{\rm Dir.}(\vec{r}) + \mathcal{E}_{\rm Coul}^{\rm Exc.}(\vec{r}),
\end{equation}
where the direct term is given by
\begin{equation}
\label{could}
\mathcal{E}_{\rm Coul}^{\rm Dir.}(\vec{r}) = \frac{e^2}{2}\int d^3r' \frac{\rho_p(\vec{r})\rho_p(\vec{r}')}{|\vec{r}-\vec{r}'|},
\end{equation}
and the exchange term is treated in the Slater approximation as
\begin{equation}
\label{coule}
\mathcal{E}_{\rm Coul}^{\rm Exc.}(\vec{r}) = -\frac{3e^2}{4} \qty(\frac{3}{\pi})^{1/3} \rho_p^{4/3}(\vec{r}).
\end{equation}
The volume-type pairing energy density is
\begin{eqnarray}
\label{epair}
\mathcal{E}_{\rm pair}(\vec{r}) = \sum_q \frac{V^q_0}{4}\tilde{\rho}^2_q(\vec{r}).
\end{eqnarray}

The various local densities are constructed using the quasi-particle wave 
functions that are solutions of the HFB equation~\cite{ring80,doba84} 
\begin{equation}
\label{HFBeq}
\mqty(h_{\uparrow\uparrow}-\lambda  & h_{\uparrow\downarrow}           & \tilde{h}_{\uparrow\uparrow}   & \tilde{h}_{\uparrow\downarrow}   \\
      h_{\downarrow\uparrow}        & h_{\downarrow\downarrow}-\lambda & \tilde{h}_{\downarrow\uparrow} & \tilde{h}_{\downarrow\downarrow}  \\
      \tilde{h}_{\uparrow\uparrow}   & \tilde{h}_{\uparrow\downarrow}  &  -h_{\uparrow\uparrow}+\lambda & -h_{\uparrow\downarrow}           \\
      \tilde{h}_{\downarrow\uparrow} & \tilde{h}_{\downarrow\downarrow}&  -h_{\downarrow\uparrow}       & -h_{\downarrow\downarrow}+\lambda)
\mqty(u_{\uparrow,k} \\
      u_{\downarrow,k} \\
      v_{\uparrow,k} \\
      v_{\downarrow,k})
=E_k\mqty(u_{\uparrow,k} \\
      u_{\downarrow,k} \\
      v_{\uparrow,k} \\
      v_{\downarrow,k}),
\end{equation}
where we have omitted the superscripts, $q$'s, from all quantities for
simplicity. The arrows, $\uparrow$ and $\downarrow$, denote the nucleon spin 
$\sigma=1/2$ and $\sigma=-1/2$, respectively. For the four-component
quasi-particle wave functions, we have abbreviated that 
$u_{\uparrow,k} \equiv u_k(\vec{r}\sigma=\frac{1}{2})$, for instance.

In Eq.~(\ref{HFBeq}) the particle-hole part of the mean-field Hamiltonian is 
\begin{eqnarray}
\label{mean-field}
h^q_{\sigma\sigma'}(\vec{r}) &=& 
   \qty[-\vec{\nabla} \cdot \frac{\hbar^2}{2m^*_q(\vec{r})}\vec{\nabla} + U_q(\vec{r}) + U_{\rm Coul}(\vec{r})\,\delta_{q,p}]
                        \delta_{\sigma\sigma'} \nonumber \\
   & & -\qty[i\vb{B}_q(\vec{r}) \cdot \qty(\vec{\nabla}\cross \vec{\sigma})]_{\sigma\sigma'},
\end{eqnarray}
where $\bm{\sigma}$ is a Pauli matrix.
The effective mass is defined through
\begin{equation}
\frac{\hbar^2}{2m^*_q(\vec{r})}=\frac{\hbar^2}{2m}+b_1\rho-b_1'\rho_q, 
\end{equation}
and
\begin{equation}
\vb{B}_q(\vec{r}) = b_4 \vec{\nabla}\rho(\vec{r}) + b'_4 \vec{\nabla}\rho_q(\vec{r}).
\end{equation}
The potential due to Skyrme force in Eq.~(\ref{mean-field}) reads
\begin{align}
\label{pot}
U_q(\vec{r}) &= b_0 \rho - b'_0 \rho_q + b_1 \tau - b'_1 \tau_q -b_2\nabla^2\rho + b'_2\nabla^2\rho_q \nonumber \\
    &\quad  + \frac{b_3}{3}\qty(\alpha+2)\rho^{\alpha+1}
        - \frac{b'_3}{3}\sum_q \qty(\alpha\rho^{\alpha-1} \rho^2_q+2\rho^{\alpha} \rho_q) \nonumber \\
    &\quad  -b_4 \vec{\nabla} \cdot \vec{\mathrm{J}} - b'_4\vec{\nabla}\cdot\vec{\mathrm{J}}_q.
\end{align}
For protons ($q=p$), the Coulomb potential in Eq.~(\ref{mean-field})
is composed of the direct and exchange parts
\begin{equation}
\label{coulomb}
U_{\rm Coul}(\vec{r}) = U_{\rm Coul}^{\rm Dir.}(\vec{r})+U_{\rm Coul}^{\rm Exc.}(\vec{r}),
\end{equation}
where
\begin{equation}
U_{\rm Coul}^{\rm Exc.}(\vec{r}) = -e^2\qty(\frac{3}{\pi})^{1/3} \qty[\rho_p(\vec{r})]^{1/3}.
\end{equation}
The direct part of the Coulomb potential is obtained by solving the
3D Poisson equation
\begin{equation}
\nabla^2 U_{\rm Coul}^{\rm Dir.}(\vec{r}) = - 4 \pi e^2 \rho_p(\vec{r}).
\end{equation}
For details, see Ref.~\cite{shi18}. Upon obtaining $U_{\rm Coul}^{\rm Dir.}$, 
the direct Coulomb energy density (\ref{could}) can be calculated through 
\begin{equation}
\mathcal{E}_{\rm Coul}^{\rm Dir.}(\vec{r}) = \frac{1}{2} \rho_p(\vec{r})\,U_{\rm Coul}^{\rm Dir.}(\vec{r}).
\end{equation}

The pairing mean-field Hamiltonian in Eq.~(\ref{HFBeq}) reads
\begin{equation}
\label{pair_pot}
\tilde{h}^q_{\sigma\sigma'}(\vec{r})  = 
            \frac{1}{2}V^q_0\tilde{\rho}_q(\vec{r}) \delta_{\sigma\sigma'}.
\end{equation}

The local densities appearing in
Eqs.~(\ref{EDFkinetic}), (\ref{skyrme}),
(\ref{could}), (\ref{coule}), (\ref{epair}), (\ref{mean-field}), and
(\ref{pair_pot}) are expressed in terms of the quasi-particle wave functions of
the HFB equation (\ref{HFBeq})
\begin{eqnarray}
\label{dens}
&&\rho_q(\vec{r}) = \sum_k\sum_{\sigma=\pm\frac{1}{2}} 
                  \qty|v_{k,q}(\vec{r}\vb{\sigma})|^2,\nonumber\\
\label{tau}
&&\tau_q(\vec{r}) = \sum_k\sum_{\sigma=\pm\frac{1}{2}} 
                  \qty|\bm{\nabla}v_{k,q}(\vec{r}\vb{\sigma})|^2,\nonumber\\
\label{rhot}
&&\tilde{\rho}_q(\vec{r}) = -\sum_k\sum_{\sigma=\pm\frac{1}{2}}  
v_{k,q}(\vec{r}\sigma)u_{k,q}^*(\vec{r}\sigma), \nonumber\\
\label{divj}
&&\vb{J}_q(\vec{r}) = -i\sum_k  v^{\dagger}_{k,q}(\vec{r}) \qty(\vec{\nabla} \cross \vec{\sigma}) v_{k,q}(\vec{r}).
\end{eqnarray}
When evaluating the spin-current density $\vb{J}_q$ on the right hand side of
Eq.~(\ref{divj}), we assume the form of $v_{k,q}(\vec{r}) \equiv
\mqty(v_{k,q}(\vec{r},\frac{1}{2}) \\ v_{k,q}(\vec{r},-\frac{1}{2}))$. The
summations over $k$ are limited to the states within the energy window of
$0<E_k<60$\,MeV. This is to make sure the lowest hole state is included.

In practical calculations, one starts with a set of quasi-particle wave
functions [$u_k(\vec{r}\sigma)$'s and $v_k(\vec{r}\sigma)$'s] calculated
assuming a Woods-Saxon potential in Eq. (\ref{pot}).
The pairing field (\ref{pair_pot}), which is diagonal in the $\sigma$ spcace,
is initialized with $\tilde{h}^q_{\uparrow\uparrow}(\vec{r})=\tilde{h}^q_{\downarrow\downarrow}(\vec{r})=-3.0$\,MeV
for $|\vec{r}| \le R$, and $\tilde{h}^q_{\uparrow\uparrow}(\vec{r})=\tilde{h}^q_{\downarrow\downarrow}(\vec{r})=0$ for $|\vec{r}| > R$, 
where $R=1.2 A^{1/3}$\,fm.
The obtained quasi-particle wave functions are
inserted into Eq.~(\ref{dens}) to calculated the densities. One then assembles
various potentials and obtains the HFB matrix [Eq.~(\ref{HFBeq})] in some
suitable representation. Diagonalizing it, one obtains the new
$u_k(\vec{r}\sigma)$'s and $v_k(\vec{r}\sigma)$'s. This process is continued
until convergence is achieved. To ensure stable convergence, we use linearly
mixed densities with 75\% of the densities from the last iteration.

The particle-hole [Eqs.~(\ref{pot}) and (\ref{coulomb})] and pairing
[Eq.~(\ref{pair_pot})] potentials are diagonal in the coordinate-space
representation. Their matrix forms in the HO and FD basis can be obtained
conveniently. In Appendix~\ref{den_mix}, We show how the matrix elements of
various densities are calculated in the mixed basis. The kinetic and the SO
parts in Eq.~(\ref{mean-field}) contain $\vec{\nabla}$ and $\nabla^2$
operators, which are not diagonal in the coordinate-space representation. In
Appendices~\ref{Ham_kin} and \ref{Ham_so}, we will give expressions for the
kinetic terms and SO terms in the mixed basis, respectively.

\subsection{Constraints}
\label{constraint}

For the constrained calculation, we follow the procedure in Ref.~\cite{ryss15a}.
The essential part of the constrained calculation is to add the following
contribution to the single-particle potential (\ref{mean-field})
\begin{equation}
      C\qty(\langle \hat{O}\rangle - \mu) \hat{O},
\end{equation}
where $C$ is the stiffness coefficient; $\hat{O}$ is the multipole 
operator in question;
$\mu$ is changing in each iteration with
\begin{equation}
      \mu^{(i+1)}=\mu^{(i)}-0.02\qty(\langle \hat{O}\rangle^{(i+1)} -\mu_0 ),
\end{equation}
where $\mu_0$ is the constraint value at the first iteration.  This procedure is
similar to the linear constraint method, and turns out to give fairly accurate
target constrained values at convergence.

In addition to the principal-axes constraint operators described in the next subsection, for the multipole
operators appears in this work, we use the quadrupole and octupole operators:
\begin{eqnarray}
      \hat{Q}_{20}&=&2\hat{z}^2-\hat{x}^2-\hat{y}^2, \nonumber \\
      \hat{Q}_{22}&=&\sqrt{3}\qty(\hat{x}^2-\hat{y}^2), \nonumber \\
      \hat{Q}_{30}&=& \hat{r}^3Y^*_{30}(\hat{\Omega}).
\end{eqnarray}
The $\beta_2$ and $\gamma$ deformations which measure the degree of 
axial and triaxial deformations are defined as 
\begin{eqnarray}
  \beta_2 &=&  \frac{4\pi}{3AR^2} \sqrt{ Q_{20}^2 + Q_{22}^2}, \nonumber \\
  \gamma  &=&  \atan{Q_{20}/Q_{22}}, \nonumber
\end{eqnarray}
where $R=1.2A^{1/3}$\,fm, $Q_{20}=\langle\hat{Q}_{20}\rangle$, 
and $Q_{22}=\langle \hat{Q}_{22}\rangle$.

\subsection{Numerical details}
\label{numerical}

As noted at the end of Sec.~\ref{HFB}, after solving the HFB problem in
the mixed basis, one obtains a set of eigenvalues and respective
eigenfunctions in the HO+FD mixed-basis representation, $(E_k, u^{n_x,n_y,i_z}_k, v^{n_x,n_y,i_z}_k)$, 
where we have ignored the spin and isospin for simplicity. 
The superscripts of $u$ and $v$ indicate the fact that they are vectors in the 
$|n_xn_yi_z\rangle$ space.
The eigenvectors in the 3D coordinate space can 
then be obtained using the following transform
\begin{align}
\label{transform}
u_k(i_x,i_y,i_z) =\sum_{n_x}\sum_{n_y}\psi_{n_x}(x_{i_x})\psi_{n_y}(y_{i_y})\,u^{n_x,n_y,i_z}_k, \\
v_k(i_x,i_y,i_z) =\sum_{n_x}\sum_{n_y}\psi_{n_x}(x_{i_x})\psi_{n_y}(y_{i_y})\,v^{n_x,n_y,i_z}_k,
\end{align}
where 
\begin{equation}
      \label{coordinate_values}
    \mu_{i_{\mu}}=(i_{\mu}-0.5) \times d\mu,~~\mu=x,y,z.
\end{equation}
We have used the discretized coordinates $i_x,i_y,i_z$ to mirror the situation 
in the $x$ and $y$ directions. In Eq.~(\ref{transform}),
$\psi_n(\mu)$ are 
the HO basis functions~\cite{doba97a}. We will call this representation ``grid
basis'' later.

In the implemented HFB code with the mixed basis we use two numbers to characterize
the dimension of the basis $|n_xn_yi_z\rangle$, which are $N_{\rm max}$ and
$N_z$. 
\footnote{For HO3D and \HFODD~calculations, we also use $N_{\rm max}$ to
specify the largest $N$ for the spherical HO basis $(n_x +n_y + n_z \le N_{\rm
max})$. In some~\HFODD~calculations, we use $N^{x,y,z}_{\rm max}$ to specify
the largest $N$ in the $x$-, $y$-, and $z$-direction of the non-uniform HO
basis.} 
The HO basis number is determined by requiring $n_x+n_y\le N_{\rm
max}$. The value of $N_z$ specifies the number of grid points in the $z$
direction. Using this basis, the dimension is 
$\frac{(N_{\rm max}+1)\times(N_{\rm max}+2)}{2} \times N_z\times 2$ 
for the HF calculations. For HFB calculation, the
dimension should be doubled. The spacing between grid points is given by $dz$.
It is clear that $N_{\rm max}$, $N_z$, and $dz$ uniquely specify our mixed basis.
We always select the value of $N_z$ to make sure the simulating
box is large enough so that the neutron density at the edge is smaller than
10$^{-5}$\,fm$^{-3}$. We use nine-point formulae for both the first- and second-derivative 
operators with respect to $z$ which will be detailed in Appendix~\ref{den_mix}. 
For the HO basis, we use an oscillator constant of
\begin{equation}
b_{\mu}=\sqrt{\frac{41\times0.6}{A^{\frac{1}{3}}\times20.7355}}~{\rm fm}^{-1},~~~\mu=x,y,z.
\end{equation}

To minimize the difference when comparing the results from HFB$^{\rm mix}$ with
those from \HFODD, we will readjust the pairing strengths of
\HFODD~calculations to match the pairing energy [Eq. (\ref{epair})] of
HFB$^{\rm mix}$. To check consistency of such a procedure between the two
calculations, another useful quantity is the average pairing gap, which is
defined as
\begin{equation}
\label{Delta}
\Delta_q=-\frac{1}{2N_q}\int d^3r \sum_{\sigma=\pm\frac{1}{2}} \tilde{h}^q_{\sigma\sigma}(\vec{r})\rho_q(\vec{r}),
\end{equation}
where $N_q$ denotes the proton or neutron number.

For the Coulomb potential, we solve the Poisson equation with two points
outside the simulating box as the boundary condition. In the current work, we
adopt the Dirichlet boundary condition. Using very similar choices of boundary
conditions and discretization method (FD method) in the $z$ direction, we note
a good description of spatial vibrations of nucleus~\cite{shi20}.

Ideally, one can impose $y$-simplex symmetry for the wave functions, rendering a
block diagonal structure of the HFB matrix in Eq.~(\ref{HFBeq}). However, at
large deformation, like the case that will be shown in Sec. \ref{pu240_sec} for
the fissioning $^{240}$Pu, the broken symmetries are encountered often. To
allow for future time-dependent extensions capable of complex simulations at
large deformation, the currently implemented mixed-basis HFB problem does not
impose any spatial symmetry.

If the initial quasi-particle wave functions possess certain symmetry, for
instance reflection symmetry, the iteration procedure will preserve the
symmetry (known as the self-consistent symmetry~\cite{ring80}). Due to the
self-consistent symmetry, most of our calculations which are started from a set
of wave functions conserving parity will continue to conserve parity symmetry.
We always make sure that the nucleus' principal axes coincide with the
Cartesian system by checking the expectation values of the $\hat{x}$,
$\hat{y}$, $\hat{z}$, $\hat{x}\hat{y}$, $\hat{x}\hat{z}$, and $\hat{y}\hat{z}$
operators are close to zero. For instance, the center of mass of the triaxially
deformed ground state of $^{110}$Mo deviates from the origin by only $\langle
x,y,z \rangle < 10^{-13}$\,fm at convergence. Thus, except for the calculation
of the fissioning path of $^{240}$Pu where the parity symmetry is explicitly
broken by the $\hat{Q}_{30}$-constraining term, we do not add constraints to
fix the orientation or the center of mass of the nucleus. When the axial
octupole moment ($Q_{30}$) of $^{240}$Pu deviates from zero, the dipole moment
(with operator $\hat{z}$) is constrained to be zero in order to prevent the
center of mass of the nucleus from moving around.

\section{Results and Discussions}
\label{results}

The HF results provide perhaps the most suitable testing cases for benchmarking
codes using different representations. This is because, in the HF case, the
occupied single-particle orbits, which are bound, can be made {\it identical}
for the codes using different representations, whereas in the HFB case, this
can hardly be achieved. Indeed, to include all the deep-hole states in HFB, one
has to include the continuum state in the configuration space. The description
of the quasi-particle states in the HFB depends on how the continuum state is
discretized in the employed basis representation. This point will be
illustrated in Sec.~\ref{hfb_bench}.  For instance, the HFB calculation for
$^{120}$Sn with the mixed basis includes 676 proton and 772 neutron
quasi-particle states for $0<E_k<60$\,MeV. In contrast, there are 574 proton
and 616 neutron quasi-particle states using the 3D HO basis.

\subsection{HF Results}
\label{hf}

In this subsection, we perform calculations for two spherical nuclei ($^{16}$O
and $^{208}$Pb), one prolately deformed nucleus ($^{24}$Mg), and one triaxially
deformed nucleus ($^{64}$Ge) using the HF$^{\rm mix}$ code. We compare the
results with those of the \HFODD~code to show the accuracy of the developed
mixed-basis code on the HF level. Before evaluating the FD part of the
HFB$^{\rm mix}$ code, careful comparisons are made for $^{16}$O between HO3D
and \HFODD~codes to check the accuracy of the HO part of the HFB$^{\rm mix}$
code.

\subsubsection{HO3D v.s. \HFODD}

Before constructing the HO+FD mixed-basis code, we first implement a 3D
HO-basis code, HO3D. This effort is not to have another 3D HO-basis code
similar to \HFODD, but the HO3D code should be considered as an intermediate
byproduct leading to the final mixed-basis code. In this subsection, we first
compare the results of HO3D with those of \HFODD~code. The purpose is to check
the implementation of the HO part of the mixed-basis code. In
Table~\ref{o16_3dhf}, we list the results of $^{16}$O calculated using HO3D and
\HFODD~codes in the same HO model space with $N_{\rm max}=14$.

In HO3D, we used the simple trapezoidal rule for the integration over the
coordinates in the HO basis. We did not employ the Gauss-Hermite quadrature
which is employed in \HFODD~code, because the basis in the $z$ direction of the
mixed-basis code is expressed in the equal-distance grid points, and it is more
straightforward to implement the same numerical technique for the integrals and
derivatives in $x$, $y$, and $z$ directions.

Even with the above-mentioned differeces, we see that the agreement between the
two codes on the total energy is less than $1$\,keV for $dz=0.6$\,fm. In an
earlier study~\cite{shi18}, the HO wave functions are represented in the FD
method and the integrals are calculated using the trapezoidal rule. Comparing
Table~\ref{o16_3dhf} with Table II of Ref.~\cite{shi18}, we observe a similar
convergence pattern.

\begin{table*}[htb]
\caption{The total energy of $^{16}$O and its decomposition into $E_{\rm
kin+c.m.}$, $E_{\rm Skyrme}$, and $E_{\rm Coul}$ (in MeV) calculated in
$N_{\rm max}=14$ model space using \HFODD~and HO3D with different spatial
discretizations ($dz$) for integration. The $E_{\rm Skyrme}$ is further
decomposed into $E_{\rho^2}$, $E_{\rho\tau}$, $E_{\rho^{2+\alpha}}$,
$E_{\rho\nabla^2\rho}$, and $E_{\rm SO}$ which correspond to the space integrals
of terms with coefficients ($b_0$, $b_0'$), ($b_1$, $b_1'$), ($b_3$, $b_3'$),
($b_2$, $b_2'$), and ($b_4$, $b_4'$) in Eq.~(\ref{skyrme}), respectively.
Skyrme SLy4 force is employed.}
\label{o16_3dhf}
\begin{ruledtabular}
\begin{tabular}{lrrrrr}
& \multicolumn{4}{c}{HO3D ($N_{\rm max}=14$)}  & \HFODD \\
 \cline{2-5}
 & $dz=0.9$\,fm & $dz=0.833$\,fm & $dz=0.75$\,fm & $dz=0.6$\,fm & ($N_{\rm max}=14$) \\
\hline
$E_{\rm Total}$               & $-$128.40483 & $-$128.46119   & $-$128.46798  & $-$128.46{\bf 573}   & $-$128.46{\bf 614}   \\
$E_{\rm kin+c.m.}$            &  222.23732  & 222.16985      & 222.18856    &    222.1{\bf 7432}    & 222.1{\bf 8044}       \\
$E_{\rho^2}$                  & $-$1304.7338 & $-$1303.3176   & $-$1303.4058 & $-$1303.3{\bf 125} & $-$1303.3{\bf 417}    \\
$E_{\rho\tau}$                &   50.50790   & 50.45808        & 50.46788    &     50.46{\bf 340}    & 50.46{\bf 476}      \\
$E_{\rho^{2+\alpha}}$         &  829.76110  & 828.75045       & 828.81217     &    828.7{\bf 4865}    & 828.7{\bf 6895}      \\
$E_{\rho\nabla^2\rho}$          & 61.18723  & 60.84091    & 60.83204            &      60.82{\bf 523}      & 60.82{\bf 684} \\
$E_{\rm SO}$                  & $-$0.95678  & $-$0.95063     & $-$0.94971    &      $-$0.949{\bf 94}    & $-$0.94962 \\
$E_{\rm Coul}^{\rm Dir.}$    &   16.40879  & 16.40333       & 16.40257      &       16.{\bf 40072}     & 16.{\bf 39990} \\
$E_{\rm Coul}^{\rm Exc.}$    & $-$2.81656   & $-$2.81562     & $-$2.81571   &     $-$2.8156{\bf 2}    & $-$2.8156{\bf 6} \\
\end{tabular}
\end{ruledtabular}
\end{table*}

\subsubsection{HF$^{\rm mix}$ v.s. HFODD: Spherical nuclei}

In Table~\ref{o16} we present a set of HF results of $^{16}$O using HF$^{\rm
mix}$ and \HFODD~codes. The model space for \HFODD~is $N_{\rm max}=11$, while
it is $N_{\rm max}=11$ for the HO part of HF$^{\rm mix}$. Five $dz$ values
(1.0, 0.9, 0.833, 0.75, and 0.6\,fm) are used to see the convergence property
of the code with decreasing grid spacing. For each grid spacings, the $N_z$
values are listed in the table. This corresponds to $z_{\rm max}=(N_z-1) \times
dz/2\approx8$\,fm. Because at the edge of the boxes the neutron or proton
densities are well below 10$^{-6}$\,fm$^{-3}$, the size of $N_z$ is not the
main factor affecting the precision of these calculations.

For the HF$^{\rm mix}$ results, we see a good convergence of the total energy
with decreasing $dz$. For $dz=0.9$\,fm, the total energy overbinds by only
$<50$\,keV compared to the result of $dz=0.6$\,fm. These mixed-basis results
show very small quadrupole moments ($Q_{20}<0.1$\,fm$^2$) due to the breaking
of the spherical symmetry in the basis.  The smallness of the quadrupole
deformation calculated for $^{16}$O can be considered as a good check for the
correctness of the code using a mixed basis.  Comparing the HF$^{\rm mix}$
results for $^{16}$O with \HFODD~ones, we notice a good agreement between them.
Indeed, the total energy calculated by the \HFODD~code differs from those of
HF$^{\rm mix}$ ($dz=0.9$, 0.833, 0.75, and 0.6\,fm) by only less than
$50$\,keV.

\begin{table*}[htb]
\caption{The total energy of $^{16}$O and its decomposition into $E_{\rm
kin+c.m.}$, various Skyrme SLy4 energies (defined in Table~\ref{o16_3dhf}), and
$E_{\rm Coul}$ as well as the quadrupole moments calculated using
 HF$^{\rm mix}$ ($N_{\rm max}=11$) with different
spatial discretizations ($dz$) and \HFODD~($N_{\rm max}=11$). All values
are in MeV, except for $Q_{20}$ and $\beta_2$ which are in fm$^2$ and dimensionless.}
\label{o16}
\begin{ruledtabular}
\begin{tabular}{lrrrrrr}
& \multicolumn{5}{c}{HF$^{\rm mix}$($N_{\rm max}=11$)}  & \HFODD \\
 \cline{2-6}
 & $dz=1.0$\,fm & $dz=0.9$\,fm & $dz=0.833$\,fm & $dz=0.75$\,fm & $dz=0.6$\,fm & $N_{\rm max}=11$ \\
 & $N_z=18$     & $N_z=18$     & $N_z=22$         & $N_z=22$        & $N_z=30$  \\
\hline
$E_{\rm Total}$                 & $-$128.570 & $-$128.504 & $-$128.490   & $-$128.482 & $-$128.476 & $-$128.450\\
$E_{\rm kin+c.m.}$              &  222.523  & 222.315    & 222.276      & 222.236    &  222.199   & 222.348 \\
$E_{\rho^2}$                  & $-$1305.807 & $-$1304.498 & $-$1303.864   & $-$1303.577 & $-$1303.272 & $-$1303.658 \\
$E_{\rho\tau}$                &  50.614  & 50.545      & 50.498        & 50.478      & 50.456     & $-$364.383 \\
$E_{\rho^{2+\alpha}}$         & 830.487 & 829.602    & 829.127       & 828.920 & 828.701 & 828.944 \\
$E_{\rho\nabla^2\rho}$          & 60.942 & 60.880  & 60.825    & 60.819   & 60.805 & 60.809 \\
$E_{\rm SO}$                  & $-$0.931 & $-$0.942  & $-$0.942     & $-$0.946    & $-$0.949 & $-$0.940 \\
$E_{\rm Coul}^{\rm Dir.}$    &  16.418  & 16.410     & 16.406       & 16.403      &   16.400  & 16.404 \\
$E_{\rm Coul}^{\rm Exc.}$    & $-$2.818   & $-$2.817   & $-$2.816     & $-$2.816   & $-$2.816   & $-$2.817 \\
$Q_{20}$                      &   0.0238  & 0.0417     & 0.0315       & 0.0713      & 0.0889 & 0.0000 \\
$\beta_2$                     &   0.0002  & 0.0004     & 0.0003       & 0.0006      & 0.0008 & 0.0000 \\
\end{tabular}
\end{ruledtabular}
\end{table*}

Table~\ref{pb208} lists the results for the heaviest doubly magic nucleus
$^{208}$Pb using the HF$^{\rm mix}$ and the \HFODD~codes. For the HF$^{\rm
mix}$ calculations, the integration is performed over the cubic box
($\approx[-12,+12]^3$\,fm$^3$). This model space is large enough that the
densities at the edges are smaller than 10$^{-6}$\,fm$^{-3}$. The results are
shown for $dz=0.9$ and $0.833$\,fm with $N_{\rm max}=12$ and 14. 
One can see that the HF$^{\rm mix}$ calculation with $dz=0.9$\,fm provides
reasonably converged results, with the energy differences between the two grid
spacings being $\approx50$\,keV. The total energy calculated with the HF$^{\rm
mix}$ is approximately 3 MeV (1 MeV) more bound for $N_{\rm max}=12$ $(14)$
than that obtained with the \HFODD. This is due to the limited number of HO
bases used. It can be expected that both codes would give closer results when
$N_{\rm max}$ is increased, which is similar to the case of $^{16}$O in
Table~\ref{o16}.

\begin{table*}[htb]
\caption{Similar to Table~\ref{o16}, but for $^{208}$Pb.
}
\label{pb208}
\begin{ruledtabular}
\begin{tabular}{lrrrrrr}
& \multicolumn{2}{c}{HF$^{\rm mix}$($N_{\rm max}=12$)} & \HFODD & \multicolumn{2}{c}{HF$^{\rm mix}$($N_{\rm max}=14$)} & \HFODD \\
 \cline{2-3}\cline{5-6}
 & $dz=0.9$\,fm & $dz=0.833$\,fm & $N_{\rm max}=12$ & $dz=0.9$\,fm & $dz=0.833$\,fm & $N_{\rm max}=14$ \\
 & $N_z=26$  & $N_z=30$  &  & $N_z=26$  & $N_z=30$  & \\
\hline
$E_{\rm Total}$               & $-$1633.868 & $-$1633.816  & $-$1630.639  & $-$1635.052 & $-$1634.979 & $-$1634.148  \\
$E_{\rm kin+c.m.}$            & 3864.117     &  3863.738  &  3861.319 &  3864.062  & 3863.647 &  3860.845 \\
$E_{\rho^2}$                  & $-$22378.402   & $-$22375.298 & $-$22359.370  & $-$22376.794 & $-$22373.432 & $-$22359.968\\
$E_{\rho\tau}$                &  1330.876      & 1330.586  &    1331.301   &  1329.914    &  1329.599  &  1328.865\\
$E_{\rho^{2+\alpha}}$         & 14534.801      & 14532.581 & 14523.995 & 14532.424 & 14530.027 & 14521.523 \\
$E_{\rho\nabla^2\rho}$          & 314.241        & 314.176  &   311.518  &  314.988  & 314.921 & 314.602 \\
$E_{\rm SO}$                  & $-$96.053      & $-$96.126 &  $-$95.774  & $-$96.174    & $-$96.247 & $-$96.375 \\
$E_{\rm Coul}^{\rm Dir.}$     & 827.807      &  827.783  &   827.606  &  827.789   & 827.765 &  827.608 \\
$E_{\rm Coul}^{\rm Exc.}$     & $-$31.256    & $-$31.255  &  $-$31.233  & $-$31.260   & $-$31.259 & $-$31.248 \\
$Q_{20}$                      & 6            &    8   & 0       &     $-$3     & $-$2 & 0 \\
$\beta_2$                     & 0.001        &    0.001 & 0.000   &     0.000 & 0.000 & 0.000 \\
\end{tabular}
\end{ruledtabular}
\end{table*}

\begin{figure}[htbp]
\centering
\includegraphics[width=0.45\columnwidth]{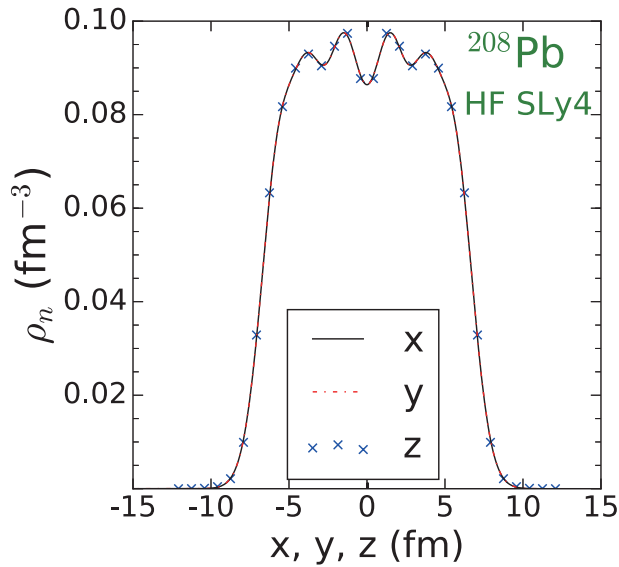}
\caption{The neutron density profiles along $x$, $y$, and $z$ directions for 
$^{208}$Pb. The density curves of $\rho_n(x,a,a)$, $\rho_n(a,y,a)$, and the 
  discrete points $\rho_n(a,a,z)$ are shown, where $a=0.45$ fm.
}
\label{figure1}
\end{figure}

Figure~\ref{figure1} plots the neutron density profiles along the three
Cartesian directions for the cross sections that are closest to the center of
the $^{208}$Pb nucleus ($dz=0.45$\,fm, $N_{\rm max}=12$). For instance, the $x$
profile denotes the neutron density with $y=z=0.45$\,fm
\nolinebreak($dz=0.9$\,fm), which is the closest point to the center of the
nucleus included in the discretization. For this heavy nucleus, one notices
rather small difference (The largest difference appears at $z=1.35$\,fm,
$\Delta\rho\approx0.0004$\,fm$^{-3}$.) between the density profile along $z$
direction and those along the $x$ and $y$ directions. The difference is
probably because the density is still converging with increasing $N_{\rm max}$
in the $x$ and $y$ directions.

\begin{figure}
\centering
\includegraphics[width=0.45\columnwidth]{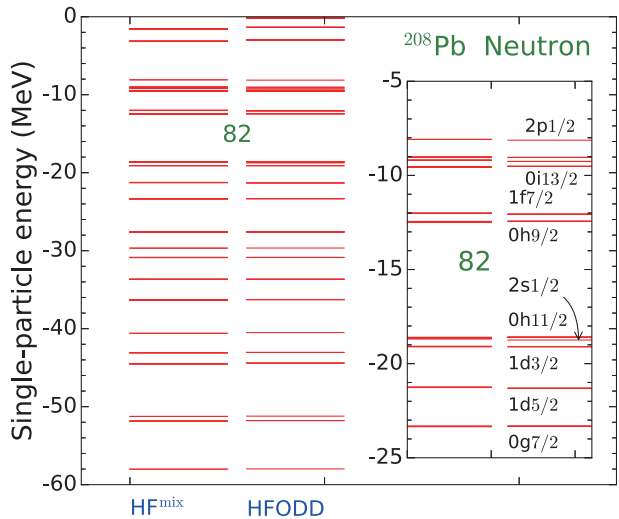}
\caption{The single-neutron energies for $^{208}$Pb calculated with HF$^{\rm mix}$ 
and \HFODD~codes.}
\label{figure2}
\end{figure}

In Figure~\ref{figure2}, we plot the single-neutron energies for $^{208}$Pb
using HF$^{\rm mix}$ ($dz=0.9$\,fm, $N_{\rm max}=12$) and compare them with the
results using the \HFODD~code ($N_{\rm max}=12$). We notice good agreement
between the two codes. For the HF$^{\rm mix}$ result we see small energy
splittings for the single-neutron energies due to the breaking of the spherical
symmetry in the basis, whereas the spherical results of \HFODD~show perfect
degeneracy. For the $0h_{9/2}$ levels, the energy splitting in HF$^{\rm mix}$
is about 0.03 MeV. Even in the zoomed-in plot shown in the inset of
Fig.~\ref{figure2}, the widths are negligible compared to the large gaps
which are of a few MeV.

\subsubsection{Prolate nucleus: $^{24}$Mg}

In Table~\ref{mg24}, we present a set of calculations for a light deformed
nucleus $^{24}$Mg performed for fixed $(N_{\rm max}, dz, N_z$) = (13,
0.833\,fm, 22) and (13, 0.9\,fm, 18), with $z$ and $x$ being the symmetry axes.
It can be seen that the results for both variants agree on the level of a few
tens of keV ($\approx30$\,keV). Decreasing $dz$ to 0.833\,fm results in the
absolute binding energies a few tens of keV smaller than those corresponding to
$dz=0.9$\,fm. Comparing the HF$^{\rm mix}$ results ($N_{\rm max},dz$)=(13,
0.833\,fm) with those of \HFODD, we see the agreement is on the order of tens
of keV. This better agreement compared to $^{208}$Pb might be because the
energies have been saturated to the basis-number parameters, $N_{\rm max}$ and
$dz$, for this lighter nucleus. It is satisfying to note that the two codes
agree on the calculated $\beta_2$ values up to the first two significant
digits.

\begin{table*}[htb]
\caption{The calculated energies (in MeV), quadrupole moments (in fm$^2$), 
and deformation parameters for $^{24}$Mg using the HF$^{\rm mix}$ and \HFODD~codes.
We include results with the symmetry axis of the nucleus being aligned along the 
$z$ and $x$ directions.}
\label{mg24}
\begin{ruledtabular}
\begin{tabular}{lrrrrr}
  & \multicolumn{2}{c}{$N_{\rm max}=13$, $dz=0.9$\,fm}  & \multicolumn{2}{c}{$N_{\rm max}=13$, $dz=0.833$\,fm} & \HFODD  \\
  &  \multicolumn{2}{c}{$N_z=18$} & \multicolumn{2}{c}{$N_z=22$} & $N_{\rm max}=13$ \\ 
\cline{2-3}\cline{4-5}
 & along $z$ & along $x$ & along $z$ & along $x$ &  \\
\hline
$E_{\rm Total}$               & $-$195.732 & $-$195.698 & $-$195.670 & $-$195.701 & $-$195.660\\
$E_{\rm kin+c.m.}$            & 380.5068  & 380.539 & 380.757 & 380.582 & 380.462   \\
$E_{\rho^2}$                  & $-$2164.306 & $-$2164.325 & $-$2163.843  & $-$2163.843 &  $-$2161.812 \\
$E_{\rho\tau}$                & 97.389 & 97.397  &  97.374   &  97.352  &  97.195\\
$E_{\rho^{2+\alpha}}$         & 1395.482 &  1395.493   &   1395.401    & 1395.109 &  1393.607\\
$E_{\rho\nabla^2\rho}$          & 89.039 & 89.008  &  88.960  &  88.946  & 88.804 \\
$E_{\rm SO}$                  & $-$22.369 & $-$22.335  &  $-$22.418   &   $-$22.370  & $-$22.389 \\
$E_{\rm Coul}^{\rm Dir.}$    & 32.913 & 32.914 &  32.919 & 32.912 & 32.903  \\
$E_{\rm Coul}^{\rm Exc.}$    & $-$4.388 & $-$4.388 & $-$4.389 & $-$4.388 & $-$4.387  \\
$Q_{20}$; $Q_{22}$            & 112; 0 & $-$56; 97 & 111; 0 & $-$56; 96 & 112; 0  \\
$\beta_2$; $\gamma$           & 0.515; 0$^{\circ}$  & 0.515; 120$^{\circ}$ & 0.510; 0$^{\circ}$  & 0.511; 120$^{\circ}$ & 0.515; 0$^{\circ}$  \\
\end{tabular}
\end{ruledtabular}
\end{table*}

\subsubsection{Triaxial nucleus: $^{64}$Ge}

Finally, in Table~\ref{ge64} we list the calculated energies and quadrupole
moments for the triaxially deformed ground states of $^{64}$Ge, using the
HF$^{\rm mix}$ and \HFODD~codes. For both variants, we use $N_{\rm max}=10$ and
12. For fixed $N_{\rm max}$, decreasing $dz$ from 0.9\,fm to 0.83\,fm results
in a $\approx40$\,keV less absolute binding energy. For $N_{\rm max}=10$, the
total energy of HF$^{\rm mix}$ is $\approx1.2$\,MeV more bound compared to that
of \HFODD~($N_{\rm max}=10$). Increasing $N_{\rm max}$ to 12 results in the
difference of total energies between HF$^{\rm mix}$ and \HFODD~being only
$\approx0.5$\,MeV. The difference is expected to decrease with increasing
$N_{\rm max}$. Again, the calculated $\beta_2$ and $\gamma$ for this
medium-heavy triaxial nucleus, are almost identical for all the variants of the
calculations.

\begin{table*}[htb]
\caption{Similar to Table~\ref{pb208}, but for $^{64}$Ge.}
\label{ge64}
\begin{ruledtabular}
\begin{tabular}{lrrrrrr}
& \multicolumn{2}{c}{HF$^{\rm mix}$($N_{\rm max}=10$)} & \HFODD &\multicolumn{2}{c}{HF$^{\rm mix}$($N_{\rm max}=12$)} & \HFODD  \\
 \cline{2-3}\cline{5-6}
 & $dz=0.9$\,fm & $dz=0.83$\,fm & $N_{\rm max}=10$ & $dz=0.9$\,fm & $dz=0.83$\,fm & $N_{\rm max}=12$\\
 & $N_z=22$ & $N_z=26$ &  & $N_z=22$ & $N_z=26$ & \\
\hline
$E_{\rm Total}$               & $-$542.795   &  $-$542.756   & $-$541.610 & $-$543.086 & $-$543.045 & $-$542.564 \\
$E_{\rm kin+c.m.}$            & 1107.085     &  1106.908  &  1107.583 & 1107.070 & 1106.877 & 1107.151 \\
$E_{\rho^2}$                  & $-$6571.053  &  $-$6569.677  &  $-$6568.314& $-$6569.662 & $-$6568.087 & $-$6570.121 \\
$E_{\rho\tau}$                & 339.740      &  339.635  & 339.847 & 339.471 & 339.354 & 339.788 \\
$E_{\rho^{2+\alpha}}$         & 4296.263     &  4295.267  & 4294.664 & 4294.943 & 4293.799 &  4295.701\\
$E_{\rho\nabla^2\rho}$          & 159.256      &  159.221  & 158.793 & 159.143 & 159.089 & 159.172 \\
$E_{\rm SO}$                  & $-$39.230    &  $-$39.234  & $-$39.387 & $-$39.176 & $-$39.182 & $-$39.411 \\
$E_{\rm Coul}^{\rm Dir.}$     & 177.485      &  177.465  & 177.550 & 177.465 & 177.444 & 177.499 \\
$E_{\rm Coul}^{\rm Exc.}$     & $-$12.342    &  $-$12.340  & $-$12.347 & $-$12.341 & $-$12.339 & $-$12.342 \\
$Q_{20}$; $Q_{22}$             & 259; 135      &  259; 135  & 258; 135 & 258; 136  & 258; 136 & 255; 135 \\
$\beta_2$; $\gamma$             & 0.262; 28$^{\circ}$  &  0.262; 28$^{\circ}$  & 0.261; 28$^{\circ}$ & 0.261; 28$^{\circ}$  &  0.261; 28$^{\circ}$ &  0.259; 28$^{\circ}$ \\
\end{tabular}
\end{ruledtabular}
\end{table*}

\subsection{HFB Results}
\label{hfb_bench}

In this section, we present the calculated HFB results for one spherical
nucleus ($^{120}$Sn), one prolately deformed nucleus ($^{34}$Mg), one
triaxially deformed nucleus ($^{110}$Mo) using the HFB$^{\rm mix}$ code. We
compare them with those calculated using the \HFODD~code. In addition, we
include HFB$^{\rm mix}$ results for the typical configurations on the fission
path of $^{240}$Pu.

\subsubsection{Spherical nucleus: $^{120}$Sn}

Table~\ref{sn120} lists the results of the HFB calculations for $^{120}$Sn
using the HFB$^{\rm mix}$ and \HFODD~codes. The same $N_{\rm max}$ is used for
both calculations. The neutron pairing strength $V_0^n$ is set to $-200.0$ MeV
fm$^3$. In \HFODD, the results with a reduced neutron pairing strength that
provides a similar neutron pairing energy to HFB$^{\rm mix}$ are also shown.
The total energy for $^{120}$Sn of HFB$^{\rm mix}$ is $\approx2$\,MeV lower
than that of the \HFODD~calculation. This may be due to the relatively small
$N_{\rm max}$ (and large $dz$) used and the different ways of discretizing the
continuum for HFB$^{\rm mix}$ and \HFODD.

\begin{table}[htb]

\caption{The calculated energies, Fermi energies, and pairing gaps (in MeV),
quadrupole moments (in fm$^2$), and $\beta_2$ for $^{120}$Sn using the
HFB$^{\rm mix}$ and \HFODD~codes. The Skyrme SLy4 force is used. The neutron
pairing strengths are also listed in the first row, which are in MeV\,fm$^3$.}
\label{sn120}
\begin{ruledtabular}
\begin{tabular}{lrrrr}
 & \multicolumn{2}{c}{HFB$^{\rm mix}$($N_{\rm max}=10$)} & \multicolumn{2}{c}{\HFODD} \\
\cline{2-3} \cline{4-5}
& $dz=1.0$\,fm & $dz=0.9$\,fm & \multicolumn{2}{c}{$N_{\rm max}=10$}  \\
& $N_z=22$  & $N_z=22$ & \\
\hline
$V_0^n$                         & $-$200.0    & $-$200.0    & $-$194.8    & $-$200.0 \\
$E_{\rm Total}$                 & $-$1017.506 & $-$1017.398 & $-$1015.458 & $-$1015.818 \\
$E_{\rm kin+c.m.}$              & 2174.025    & 2173.8967    & 2170.700    & 2174.099   \\
$E_{\rho^2}$                  & $-$12683.665 & $-$12680.942 & $-$12655.007  & $-$12661.993  \\
$E_{\rho\tau}$                & 717.670 & 717.416  &  716.451   &  717.525 \\
$E_{\rho^{2+\alpha}}$         & 8264.085 &  8261.9656   &   8244.251   & 8249.274 \\
$E_{\rho\nabla^2\rho}$          & 224.724 & 224.619  &  222.092  &  222.655 \\
$E_{\rm SO}$                  & $-$49.251 & $-$49.284  &  $-$48.554   &   $-$49.914  \\
$E_{\rm Coul}^{\rm Dir.}$    & 366.623     & 366.592     & 366.375     & 366.414   \\
$E_{\rm Coul}^{\rm Exc.}$    & $-$19.104   & $-$19.103   & $-$19.083   & $-$19.085  \\
$E^{\rm n}_{\rm pair}$          & $-$12.614   & $-$12.559   & $-$12.682   & $-$14.793  \\
$\lambda_{\rm n}$               & $-$7.974    & $-$7.972    & $-$7.959    & $-$7.951  \\
$\Delta_{\rm n}$    & 1.436      & 1.433      & 1.434       & 1.572 \\
$Q_{20}$                        & 7          & $-$1           & 0           & 0  \\
$\beta_2$                       & 0.002      & $-$0.0003      & 0.0         & 0.0  \\
\end{tabular}
\end{ruledtabular}
\end{table}

\begin{figure}[htbp]
\centering
\includegraphics[width=0.55\columnwidth]{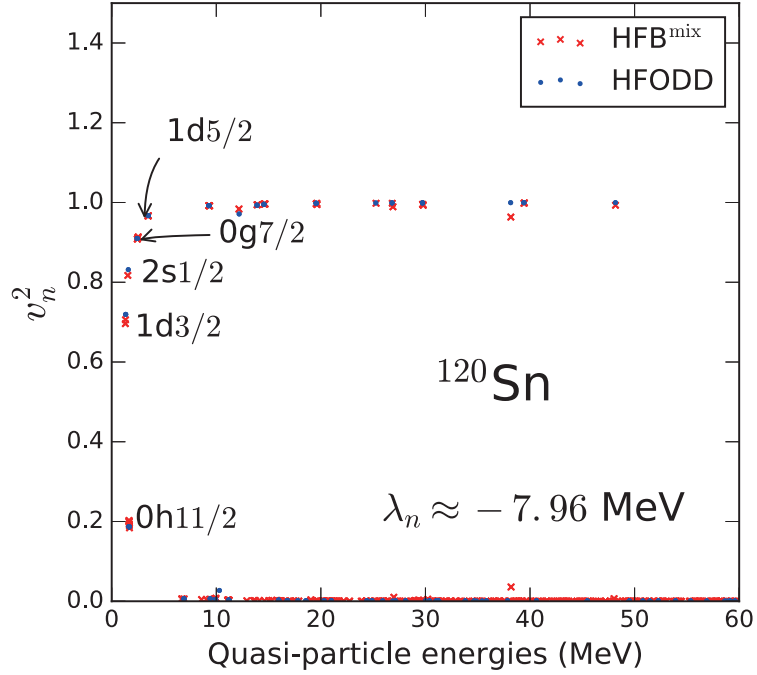}
\caption{The calculated occupation probabilities $v^2_n$'s against the 
quasi-particle energies for $^{120}$Sn.}
\label{figure3}
\end{figure}

Figure~\ref{figure3} plots the occupation probabilities
\begin{equation}
v_{k,q}^{2}=\sum_{\sigma=\pm\frac{1}{2}} \int d^3r|v_k^q(\vec{r},\sigma)|^2,
\label{v2}
\end{equation}
as a function of the quasi-particle energies for $^{120}$Sn, calculated with
HFB$^{\rm mix}$ ($dz=1.0$\,fm) and \HFODD~codes ($V_0^n=-194.8$\,MeV\,fm$^3$).
We see a nice agreement between the two codes, especially for the occupied
deep-hole states. A small deviation of $E_k$ and $v^2_k$ from a perfect
degeneracy of the HFB$^{\rm mix}$ calculation can be seen, which is associated
with the breaking of the spherical symmetry with the small quadrupole
deformation, as can be assessed from Table~\ref{sn120}. For those states with
$v^2_n$'s close to zero, we see the HFB$^{\rm mix}$ gives much more states
inside the $E_{k}<60$\,MeV window than the \HFODD~does. This is expected
because the two codes discretize the continua for $E_{k}>-\lambda$ differently.

\subsubsection{Prolate nucleus: $^{34}$Mg}

In Table~\ref{mg34}, we list the results of the HFB calculations with HFB$^{\rm
mix}$ ($N_{\rm max}=14$, $dz=1.0$\,fm) for $^{34}$Mg and compare the results
with those of the \HFODD~code. For the \HFODD~results we include one column
with a deformed HO basis using $(N_{\rm max}^{x,y},N_{\rm max}^{z})=(13,16)$.
The last column shows the \HFODD~results with spherical HO basis $N_{\rm
max}=14$. For the two \HFODD~results, the pairing strengths have been adjusted
to match the $E_{\rm pair}^{\rm {n,p}}$ of HFB$^{\rm mix}$
calculations.

We see that the total energy calculated using HFB$^{\rm mix}$ are lower by
$<100$\,keV compared to those of \HFODD. The quadrupole moments are rather
close to each other between the three variants, with the differences being
$\approx3\%$. It has to be noted that the ground-state quadrupole moment of
this nucleus depends on the pairing strengths sensitively. The good agreement
between the three variants is thus rather satisfactory. As the pairing
strengths of the \HFODD~results are readjusted to give similar $E^{\rm
n,p}_{\rm pair}$, the extracted averaged pairing gaps are also close between
the two codes.

\begin{table}[htb]
\caption{The calculated energies (in MeV), pairing gaps (in MeV), and quadrupole 
moments (in fm$^2$) together with the pairing strengths (in MeV fm$^3$)
for $^{34}$Mg using HFB$^{\rm mix}$ and \HFODD~codes.}
\label{mg34}
\begin{ruledtabular}
\begin{tabular}{lrrr}
& HFB$^{\rm mix}$ ($N_{\rm max}=14$) & \multicolumn{2}{c}{\HFODD}  \\
\cline{3-4}
& $dz=1.0$\,fm, $N_z=18$& $(N_{\rm max}^{x,y},N_{\rm max}^{z})=(13,16)$ & $N_{\rm max}^{x,y,z}=14$ \\
\hline
$V_0^n$; $V_0^p$            & $-$218.5; $-$218.5 & $-$217.5; $-$204.0 & $-$218.5; $-$205.0 \\
$E_{\rm Total}$          & $-$257.531        & $-$257.471        & $-$257.479 \\
$E_{\rm kin+c.m.}$       & 557.036           & 557.070           & 556.983   \\
$E_{\rm Skyrme}$         & $-$835.730        & $-$835.681        & $-$835.614  \\
$E_{\rm Coul}^{\rm Dir.}$    & 31.752       & 31.733            & 31.733   \\
$E_{\rm Coul}^{\rm Exc.}$    & $-$4.217      & $-$4.217          & $-$4.217  \\
$E^{\rm n}_{\rm pair}$   & $-$3.870          & $-$3.905          & $-$3.883  \\
$E^{\rm p}_{\rm pair}$   & $-$2.501          & $-$2.472          & $-$2.480  \\
$\lambda_{\rm n}$        & $-$3.221          & $-$3.258          & $-$3.260  \\
$\lambda_{\rm p}$        & $-$20.126         & $-$20.121         & $-$20.112  \\
$\Delta_{\rm n}$    & 1.325    & 1.328             & 1.328 \\
$\Delta_{\rm p}$    & 1.262    & 1.213             & 1.215 \\
$Q_{20}$                 & 99               & 102               & 102  \\
$\beta_2$                 & 0.255          & 0.262             & 0.262\\
\end{tabular}
\end{ruledtabular}
\end{table}

\subsubsection{Triaxial nucleus: $^{110}$Mo}

Table~\ref{mo110} lists the calculated ground states of $^{110}$Mo using the
HFB$^{\rm mix}$ and \HFODD~codes ($N_{\rm max}=9$) with SkM*
force~\cite{bart82}. For HFB$^{\rm mix}$, we included results with two
orientations where the longest axes of the nucleus coincide with the $z$
($z>x>y$) and $x$ ($x>y>z$) axes. For these HFB$^{\rm mix}$ runs, the dimension
of the HFB matrix [Eq. (\ref{HFBeq})] is 4840. Each iteration takes about
20\,min to finish on Intel(R) Xeon(R) CPU E5-2697 v4 processor.

The total energy obtained with $z>x>y$ is about 0.6\,MeV lower than the result
with the longest axis in the $x$ direction. This is because the $z$ direction
is treated with the FD method allowing for better description of the spatially
extended density compared to the HO basis. It is very interesting to note that
the $\beta_2$ values of all the calculated triaxial ground states differ only
from the third significant digits.

\begin{table*}[htb]
\caption{The calculated energies, pairing gaps (in MeV), and total quadrupole
moments (in fm$^2$) of $^{110}$Mo calculated with HFB$^{\rm mix}$ and
\HFODD~codes. The Skyrme force used is SkM*. The neutron pairing strengths are
also shown in unit of MeV\nolinebreak\,fm$^3$, and for the proton pairing
strength we use $V_0^p=-170.0$\,MeV\,fm$^{3}$. The calculated nucleus is placed
in the principal axis, where $z>x>y$ indicates that the longest, medium, and
shortest axes of the nucleus are aligned in the $z$, $x$, and $y$ directions,
respectively. The $\beta_2$ and $\gamma$ deformations are defined in
Sec~\ref{constraint}.}
\label{mo110}
\begin{ruledtabular}
\begin{tabular}{lrrr}
 & \multicolumn{2}{c}{HFB$^{\rm mix}$($N_{\rm max}=9$, $dz=1.0$\,fm, $N_z=22$)} & \HFODD \\
\cline{2-3} 
 & $z>x>y$ &  $x>y>z$ & $N_{\rm max}=9$  \\
\hline
$V_0^n$ & $-$170.0 & $-$170.0 & $-$172.7 \\
$E_{\rm Total}$     & $-$920.571  & $-$919.965 & $-$918.532 \\
$E_{\rm kin+c.m.}$   & 2010.785  & 2011.839 & 2008.9216 \\
$E_{\rho^2}$         & $-$12268.080  & $-$12265.004 & $-$12235.894  \\
$E_{\rho\tau}$       & 368.900   & 368.671 & 367.849    \\
$E_{\rho^{2+\alpha}}$ & 8600.508    & 8597.598 &  8575.708     \\
$E_{\rho\nabla^2\rho}$  & 193.730   & 192.776 &  190.496  \\
$E_{\rm SO}$           & $-$73.691   & $-$73.000 & $-$72.646    \\
$E_{\rm Coul}^{\rm Dir.}$    & 266.902  & 266.847 & 266.618  \\
$E_{\rm Coul}^{\rm Exc.}$    & $-$15.749  & $-$15.740 & $-$15.718  \\
$E_{\rm pair}^{\rm n}$    & $-$3.875  & $-$3.885 & $-$3.867   \\
$\lambda_{\rm n}$    & $-$5.414  & $-$5.407 & $-$5.379  \\
$\Delta_{\rm n}$  & 0.762  & 0.763 & 0.771  \\
$Q_{20}$; $Q_{22}$            & 949; 361  & $-$780; 632 & 943; 365  \\
$\beta_2$; $\gamma$           & 0.369; 21$^{\circ}$  & 0.364; $-$39$^{\circ}$ & 0.367; 21$^{\circ}$  \\
\end{tabular}
\end{ruledtabular}
\end{table*}

\subsubsection{Selected Points on the Fission pathway of $^{240}$Pu}
\label{pu240_sec}

Finally, in Table~\ref{pu240}, we present the calculated results for the
fission pathway of $^{240}$Pu using HFB$^{\rm mix}$ (SkM*). For this
calculation, we use $(N_{\rm max}, dz, N_z)$=(8, 1.2\,fm, 26). The neutron and
proton pairing strengths are $V_0^n=V_0^p=-170.0$\,MeV\,fm$^3$. These pairing
strengths give reasonable pairing energy for the ground state of $^{240}$Pu. We
select several configurations along the fission pathway, that is, the results
corresponding to the ground state (g.s), the fission isomer (Fis. Iso.), the
first/inner fission barrier (1st Fis. Bar.), the second/outer fission barrier
(2nd Fis. Bar.), and one deformation point beyond the second fission barrier at
$Q_{20}=150$\,b. 
Here, we have assumed that the fission pathway is described in the space
spanned by the several multipole deformations. Ideally, one should obtain the
solutions corresponding to the barriers by calculating the full potential
energy surface before the scission area. However, due to the limitation of the
computational resources, we restrict our search within the points near the two
fission barriers. For example, to find the first fission barrier, we perform
calculations with $Q_{20}$ being constrained from 50 to 70\,b at a step of
5\,b.  We do not impose axial nor reflection symmetries; we allow $Q_{22}$ and
$Q_{30}$ to vary freely when $Q_{20}$ is constrained. The important
deformations which lower the fission barrier and the information on the barrier
position in the potential energy surface in the case of SkM* functional are
known from the literature such as Ref.~\cite{schunck14}.

\begin{table}[htb]
\caption{The calculated energies (in MeV), pairing gaps (in MeV), quadrupole
moments [in barn (b)], and octupole moment (in b$^{3/2}$) for selected deformation
points on the fission path of $^{240}$Pu using HFB$^{\rm mix}$ (SkM* force).}
\label{pu240}
\begin{ruledtabular}
\begin{tabular}{lrrrrr}
 & g.s. & Fis. Iso. & 1st Fis. Bar. & 2nd Fis. Bar. & \\
\hline
$E_{\rm Total}$         & $-$1794.741     & $-$1792.300      & $-$1787.055   & $-$1788.195 & $-$1790.132 \\
$E_{\rm kin+c.m.}$      &4447.739        & 4441.681          &   4442.885   &  4427.123   &  4431.331 \\
$E_{\rm Skyrme}$        & $-$7230.115    & $-$7186.925       & $-$7192.441  & $-$7135.954   &  $-$7118.399\\
$E_{\rm Coul}^{\rm Dir.}$ & 1025.625     & 992.332          &  1006.738     &  963.173     &  939.697 \\
$E_{\rm Coul}^{\rm Exc.}$ & $-$35.687    & $-$35.603        & $-$35.620     & $-$35.537     & $-$35.540\\
$E_{\rm pair}^{\rm n}$    & $-$1.399     & $-$2.937         & $-$7.532      & $-$6.579       & $-$6.241\\
$E_{\rm pair}^{\rm p}$    & $-$0.904     & $-$0.848         & $-$1.085      & $-$0.652     & $-$0.979\\
$\lambda_{\rm n}$       &  $-$6.383      & $-$6.626         & $-$6.257      & $-$6.310    & $-$6.428\\
$\lambda_{\rm p}$       &  $-$4.484      & $-$4.715         & $-$4.574      & $-$5.155     & $-$5.180\\
$\Delta_{\rm n}$ & 0.318   & 0.463    &    0.747     & 0.687  & 0.665\\
$\Delta_{\rm p}$ & 0.264   & 0.262    &    0.297     & 0.228  & 0.251\\
$Q_{20}$               & 29.9            & 83.2     &    60.6      & 120.0  & 150.3\\
$Q_{22}$               & 0               & 0        &    8.6       & 4.9    & 3.0\\
$Q_{30}$               & 0               & 0        &    0         & 6.1   & 9.4\\
\end{tabular}
\end{ruledtabular}
\end{table}

The excitation energy of the fission isomer extracted from Table~\ref{pu240} is
2.4\,MeV; the heights of the inner and outer fission barriers are 7.7 and
6.5\,MeV, respectively. Comparing these three energies with those of
Ref.~\cite{schunck14} calculated with \HFODD, we notice that our results are
smaller than theirs by $\le0.3$\,MeV. These small deviations can be attributed
to the rather small $N_{\rm max}=8$ used in our calculations. Another source of
discrepancies is the differences in the pairing treatment (both pairing
strengths and pairing types chosen).

\begin{figure}
\centering
\includegraphics[width=0.55\columnwidth]{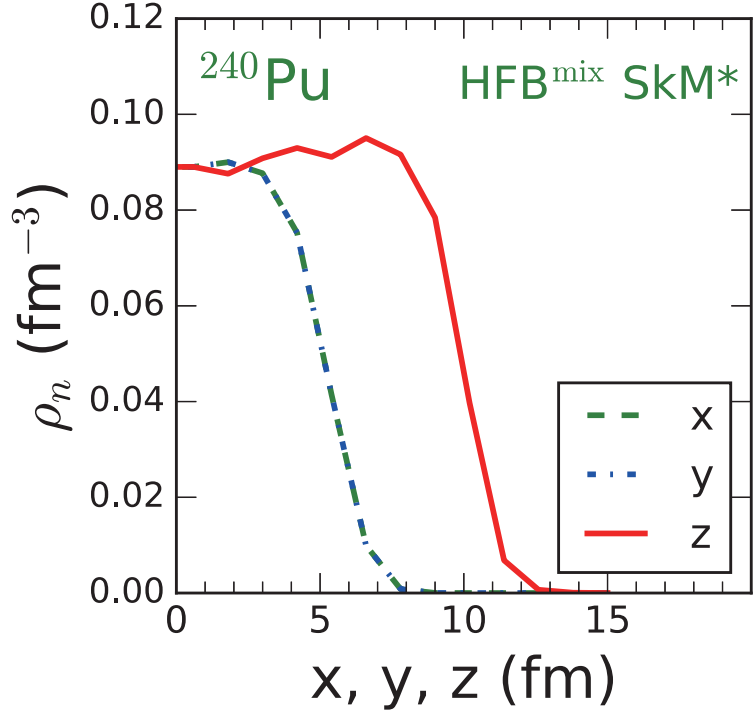}
\caption{The density profiles of the fission isomer of $^{240}$Pu calculated
using HFB$^{\rm mix}$ (Table~\ref{pu240}). The density values correspond to the
lines with other coordinates closest to the center of the nucleus.}
\label{figure6}
\end{figure}

Figure~\ref{figure6} shows the density profiles of the fission isomer along the
three axes. In Table~\ref{pu240} we include the calculated results for a
deformation beyond the second fission barrier. In the future, the HFB$^{\rm
mix}$ code will be extended to allow for the full time-dependent HFB simulation
which will be started from a HFB solution corresponding to the deformation
point beyond the second fission barrier.

Although the HFB$^{\rm mix}$ is capable of describing the extremely elongated
configuration, such as those near or beyond the scission point of the
fissioning nucleus, it is not the intention of the current paper to perform the
full fission-path calculation of $^{240}$Pu. Here we are content to demonstrate
the {\it feasibility} of the HFB$^{\rm mix}$ code for describing some of the
typical configurations of $^{240}$Pu fission pathway. We leave a more realistic
and systematic survey of the potential energy surface for the future project
with larger basis size and better pairing treatment.

\section{Summary and Perspective} 
\label{summary}

We solve the numerically demanding three-dimensional (3D) coordinate-space
nuclear Hartree-Fock-Bogoliubov (HFB) problem using a mixture of bases, which
consist of two harmonic-oscillator (HO) bases in the $x$ and $y$ directions,
and one finite-difference basis in the $z$ direction.

We implement the 3D HO and the mixed-basis codes. Using the mixed-basis code,
we perform systematic calculations for spherical and deformed nuclei using the
Hartree-Fock (HF) and HFB methods. The results obtained are compared with those
of the 3D HO-based code, \HFODD.

The differences of the total energies using the HF method for lightest nucleus
$^{16}$O is in the order of a few tens of keV between the two HO-based codes,
HO3D and \HFODD. For the lightest nuclei, $^{16}$O and $^{24}$Mg, where the
total energies tend to be converged with increasing $N_{\rm max}$, the total
energy differences are only a few tens of keV. For heavier nuclei ($A>100$),
the differences in total energies are about 1-2 MeV. The differences decrease
with the increase of the largest $N_{\rm max}$ of the HO bases. Hence, for the
heavier nuclei, the difference between mixed-basis code and \HFODD~can be
attributed to the fact that the HF result is still converging with $N_{\rm
max}$.

Comparisons of the HFB results using the HFB$^{\rm mix}$ and the \HFODD~codes
show similar convergence behavior with respect to $N_{\rm max}$ as the HF case.
Specifically, the differences of the total energies using HFB$^{\rm mix}$ and
\HFODD~decrease with increasing $N_{\rm max}$.

For spherical nuclei, the calculated quadrupole moments are negligibly small
for the mixed-basis HF and HFB codes. For deformed nuclei, the differences in
the quadrupole moments calculated using the mixed-basis codes are almost
identical to those of the \HFODD~code. A good agreement of
single/quasi-particle energies is also seen between the mixed-basis codes and
that of the \HFODD~code.

As a demonstration of the applicability of the HFB$^{\rm mix}$ code to
extremely elongated deformations, selected configurations along the fission
pathway in $^{240}$Pu are calculated. With future optimizations of the code,
the HFB$^{\rm mix}$ will be advantageous in a systematic survey of the
potential energy surfaces with elongated configurations as well as the deformed
drip-line nuclei.

Our future work will be focused on the time-dependent extension of the current
mixed-basis HFB method. Such a mixed-basis strategy would make the full 3D
time-dependent HFB calculations much more affordable and flexible.

\begin{acknowledgments}

The current work is supported by National Natural Science Foundation of China (Grant No. 12075068 and No. 11705038),
the Fundamental Research Funds for the Central Universities (Grant No. HIT.BRET.2021003),
JSPS KAKENHI Grant No. 16K17680 and No. 20K03964, 
and
the JSPS-NSFC Bilateral Program for the Joint Research Project on ``Nuclear mass and life for unravelling mysteries of r-process.''
YS thanks the HPC Studio at Physics Department of Harbin Institute of 
Technology for computing resources allocated through INSPUR-HPC@PHY.HIT.

\end{acknowledgments}

\appendix

\section{The densities in the mixed basis}
\label{den_mix}

In this section, we specifically provide the detailed formulae of various
densities~[Eq.~(\ref{dens})] using the mixed basis.

To compute the densities, $\tau$, $\div{\vb{J}}$,
and $\grad{\rho}$, which are needed for the construction
of the potentials and mean-field Hamiltonian, one needs the following first derivatives
\begin{eqnarray}
\label{first_derivative1}
\pdv{x}v_k(x_{i_x},y_{i_y},z_{i_z})&=&\sum_{n_x}\sum_{n_y}\psi'_{n_x}(x_{i_x})\psi_{n_y}(y_{i_y})\,v^{n_x,n_y,i_z}_k,\\
\label{first_derivative2}
\pdv{y}v_k(x_{i_x},y_{i_y},z_{i_z})&=&\sum_{n_x}\sum_{n_y}\psi_{n_x}(x_{i_x})\psi'_{n_y}(y_{i_y})\,v^{n_x,n_y,i_z}_k,\\
\label{first_derivative3}
\pdv{z}v_k(x_{i_x},y_{i_y},z_{i_z})&=&\sum_{n_x}\sum_{n_y}\psi_{n_x}(x_{i_x}) \psi_{n_y}(y_{i_y}) \sum_{d=-\lambda}^{+\lambda} m^{(1)}_d v_k^{n_x,n_y,i_z+d},
\end{eqnarray}
where the $m^{(1)}_d$ is the FD coefficients for the first derivative.  For the
nine-point FD case ($\lambda=4$), the coefficients for the first ($m^{(1)}_d$)
and second ($m^{(2)}_d$) derivatives are listed in Table~\ref{fd_weights}.
These are used throughout this work.

\begin{table}[htb]
\caption{The nine-point FD weights of the first and second derivatives.}
\label{fd_weights}
\begin{ruledtabular}
\begin{tabular}{lrrrrrrrrr}
$d$  & $-$4 & $-$3 & $-$2 & $-$1 & 0 & 1 & 2 & 3 & 4 \\
\hline
$m^{(1)}_d$         &  $\frac{1}{280}$    &   $-\frac{4}{105}$    &  $\frac{21}{105}$  & $-\frac{84}{105}$ & 0 & $\frac{84}{105}$ & $-\frac{21}{105}$  &  $\frac{4}{105}$ & $-\frac{1}{280}$\\
$m^{(2)}_d$         &  $-\frac{1}{560}$   &   $\frac{8}{315}$     &  $-\frac{1}{5}$    & $\frac{8}{5}$ & $-\frac{205}{72}$ & $\frac{8}{5}$ & $-\frac{1}{5}$ & $\frac{8}{315}$ & $-\frac{1}{560}$\\
\end{tabular}
\end{ruledtabular}
\end{table}

As for the term involving Laplacian ($\laplacian{\rho}$), one needs
$\laplacian{v_k}$. This can be done in a similar manner as
Eqs.~(\ref{first_derivative1}) - (\ref{first_derivative3}),
\begin{eqnarray}
\label{second_derivative1}
\pdv[2]{x}v_k(x_{i_x},y_{i_y},z_{i_z})&=&\sum_{n_x}\sum_{n_y}\psi''_{n_x}(x_{i_x})\psi_{n_y}(y_{i_y})\,v^{n_x,n_y,i_z}_k,\\
\label{second_derivative2}
\pdv[2]{y}v_k(x_{i_x},y_{i_y},z_{i_z})&=&\sum_{n_x}\sum_{n_y}\psi_{n_x}(x_{i_x})\psi''_{n_y}(y_{i_y})\,v^{n_x,n_y,i_z}_k,\\
\label{second_derivative3}
\pdv[2]{z}v_k(x_{i_x},y_{i_y},z_{i_z})&=&\sum_{n_x}\sum_{n_y}\psi_{n_x}(x_{i_x}) \psi_{n_y}(y_{i_y}) \sum_{d=-\lambda}^{+\lambda} m^{(2)}_d v_k^{n_x,n_y,i_z+d}.
\end{eqnarray}
In the case of 3D HO basis, the calculation of gradient and Laplacian of $v_k$
is more straightforward where the $z$ direction is calculated in a similar way
as those of the $x$ and $y$ directions.

After obtaining the gradient and the Laplacian of $v_k$, one
assembles the densities~\cite{doba97}
\begin{eqnarray}
\nonumber
\rho&=&D^{\uparrow\uparrow}_{00}+D^{\downarrow\downarrow}_{00}, \\
\nonumber
\tau&=&\sum_{\mu}(D^{\uparrow\uparrow}_{\mu\mu}+D^{\downarrow\downarrow}_{\mu\mu}),\\
\nonumber
\laplacian{\rho}&=&2\real(L^{\uparrow\uparrow}+L^{\downarrow\downarrow})+2\tau,\\
\nonumber
\div{\vb{J}}&=&-2\imaginary(D^{\uparrow\downarrow}_{23}-D^{\uparrow\downarrow}_{32})\\
\nonumber
            &&-2\real(D^{\uparrow\downarrow}_{31}-D^{\uparrow\downarrow}_{13})\\
\nonumber
             &&+2\imaginary(D^{\uparrow\uparrow}_{21}-D^{\downarrow\downarrow}_{21}),\\
\nonumber
\nabla_k{\rho} &=& 2\real(D^{\uparrow\uparrow}_{k 0}+D^{\downarrow\downarrow}_{k 0}),~~k=1,2,3,
\end{eqnarray}
where 
\begin{eqnarray}
D^{\sigma\sigma'}_{\mu\nu}&=&\sum_k[\nabla_{\mu}v_k(\vec{r}\sigma)][\nabla_{\nu}v^*_k(\vec{r}\sigma')], \\
L^{\sigma\sigma'}         &=&\sum_k v_k(\vec{r}\sigma)[\laplacian v^*_k(\vec{r}\sigma')],
\end{eqnarray}
and $\nabla_{\mu}\equiv(1,\bm{\nabla})$. Using Eqs.~(\ref{first_derivative1}) - (\ref{second_derivative3}), one
can also obtain the Laplacian of the effective mass in Eq.~(\ref{kinetic}).

\section{The kinetic terms in HO and mixed bases}
\label{Ham_kin}

In this subsection, we give the matrix elements of the kinetic term in the HO
and mixed-basis representations. 
Before showing the kinetic term in the coordinate representation, one notices
that, when applied to a function, the kinetic part of Eq.~(\ref{mean-field})
can be written as~\cite{jin17}
\begin{align}
	\label{kinetic}
	\hat{h}^{\rm kin.}\psi(\vec{r}) = -\frac{1}{2}\Bigg\{\frac{\hbar^2}{2m^*(\vec{r})}\nabla^2\psi(\vec{r})
	                                 +\nabla^2\qty[\frac{\hbar^2}{2m^*(\vec{r})}\psi(\vec{r})]
						             -\qty(\nabla^2\frac{\hbar^2}{2m^*(\vec{r})})\psi(\vec{r})\Bigg\}.
\end{align}
The matrix element of the third term of right hand side of Eq.~(\ref{kinetic}) is diagonal in the
FD basis.

\subsection{The matrix element of kinetic terms in the 1D grid basis}
\label{fd}

Next we examine the matrix form of the Laplacian operator $\nabla^2$ in the 1D
grid representation. The second derivative of a function in the 1D coordinate space is 
expressed as
\begin{equation}
	\label{laplacian}
	\dv[2]{f}{z}\Big|_{z_i}=\sum_{d=-\lambda}^{+\lambda}f(z_{i+d}) \, m^{(2)}_d \times (dz)^{-2}.
\end{equation}

Using the matrix form of the Laplacian operator, together with the third local term in
Eq.~(\ref{kinetic}), we have the kinetic matrix elements in the 1D FD representation
\begin{equation}
    \label{tij}
	T_{ij} = -\frac{1}{2}[h^{\rm 1D}(z_i)+h^{\rm 1D}(z_j)] \, \nabla^2_{ij} - \left[\nabla^2 h^{\rm 1D}(z)|_{z=z_i}\right] \delta_{ij}, 
\end{equation}
where $\nabla^2_{ij}= m_d^{(2)} \times (dz)^{-2} \delta_{j,i+d}$, $(|d|\le
\lambda)$ denotes the matrix elements for the Laplacian operator in the FD
basis. The function $h^{\rm 1D}(z)$ is the 1D conterpart of 
$\displaystyle \frac{\hbar^2}{2m^*(\vec{r})}$ in
Eq.~(\ref{kinetic}); $dz$ is the spacing in the $z$ direction.

\subsection{Matrix elements in the 3D coordinate-space HO basis}

The matrix elements in the 3D HO basis for the kinetic part (\ref{kinetic}) are in
the following form
\begin{eqnarray}
	\label{MEho}
	\langle n_x n_y n_z | \hat{h}^{\rm kin.} | n'_x n'_y n'_z \rangle &=& \int dx \int dy \int dz \,\psi^*_{n_x}(x) \psi^*_{n_y}(y) \psi^*_{n_z}(z) \hat{h}  \psi_{n'_x}(x) \psi_{n'_y}(y) \psi_{n'_z}(z) \nonumber \\
						 &=&-\frac{1}{2}\int dx \int dy \int dz\,\psi^*_{n_x}(x) \psi^*_{n_y}(y) \psi^*_{n_z}(z) \frac{\hbar^2}{2m^*(\vec{r})}\nabla^2 \qty[\psi_{n'_x}(x) \psi_{n'_y}(y) \psi_{n'_z}(z)] \nonumber \\
						 &&-\frac{1}{2}\int dx \int dy \int dz\,\psi^*_{n_x}(x) \psi^*_{n_y}(y) \psi^*_{n_z}(z) \nabla^2 \qty[\frac{\hbar^2}{2m^*(\vec{r})} \psi_{n'_x}(x) \psi_{n'_y}(y) \psi_{n'_z}(z)] \nonumber \\
						 &&+\frac{1}{2}\int dx \int dy \int dz\,\psi^*_{n_x}(x) \psi^*_{n_y}(y) \psi^*_{n_z}(z) \qty[\nabla^2 \frac{\hbar^2}{2m^*(\vec{r})}] \psi_{n'_x}(x) \psi_{n'_y}(y) \psi_{n'_z}(z),
\end{eqnarray}
where $\psi_{n_{\mu}}(\mu)$ is the HO wave functions in the $\mu=x$, $y$, and $z$. 
The form of the Cartesian-coordinate HO basis $|n_x n_y n_z\rangle$
is identical to those explained in Ref.~\cite{doba97a}.

\subsection{Matrix elements in the mixed basis}
\label{mix}

We write down the matrix elements in the mixed basis before a brief explanation
of each term:
\begin{eqnarray}
	\label{MEmix}
	\langle n_x n_y i_z | \hat{h}^{\rm kin.} | n'_x n'_y i'_z \rangle 
	& =&-\frac{1}{2}\int dx \int dy \, \psi^*_{n_x}(x) \psi^*_{n_y}(y) \qty[\frac{\hbar^2}{2m^*(x,y,z_{i_z})}] \qty(\frac{d^2}{dx^2}+\frac{d^2}{dy^2}) [\psi_{n'_x}(x) \psi_{n'_y}(y)]\delta_{i_z,i'_z} (dz)^{-2} \nonumber \\
	&&-\frac{1}{2}\int dx \int dy \, \psi^*_{n_x}(x) \psi^*_{n_y}(y)  \qty(\frac{d^2}{dx^2}+\frac{d^2}{dy^2}) \qty[\frac{\hbar^2}{2m^*(x,y,z_{i_z})} \psi_{n'_x}(x) \psi_{n'_y}(y)]\delta_{i_z,i'_z} (dz)^{-2} \nonumber \\
    &&-\frac{1}{2}\int dx \int dy \, \psi^*_{n_x}(x) \psi^*_{n_y}(y) \frac{1}{2}\qty[\frac{\hbar^2}{2m^*(x,y,z_{i_z})}+\frac{\hbar^2}{2m^*(x,y,z_{i'_z})}] [\psi_{n'_x}(x) \psi_{n'_y}(y)] m^{(2)}_{i_z-i'_z} (dz)^{-2} \nonumber \\
    &&+\frac{1}{2}\int dx \int dy \, \psi^*_{n_x}(x) \psi^*_{n_y}(y) \qty[\nabla^2 \frac{\hbar^2}{2m^*(\vec{r})}] \psi_{n'_x}(x) \psi_{n'_y}(y) \delta_{i_z,i'_z} (dz)^{-2},
\end{eqnarray}
where $|n_x n_y i_z\rangle$ is a 3D basis which uses the Cartesian coordinate HO
bases $\psi_{n_x}(x)$ and $\psi_{n_y}(y)$ in the $x$ and $y$ directions,
respectively. For the $z$ direction, we use the FD basis, for which the $\nabla^2$
and $\vec{\nabla}$ operators have been shown in Sec.~\ref{fd}.

The first two terms in the right-hand side of Eq.~(\ref{MEmix}) correspond to
the first two terms in Eq.~(\ref{kinetic}) for the $x$ and $y$ directions. In
these two terms, the matrix is diagonal in the $z$-direction, due to the local
property. The third term corresponds to the first two terms in
Eq.~(\ref{kinetic}) for the $z$ direction. It has been symmetrized in the same
way as explained in Sec.~\ref{fd}. The coefficients $m^{(2)}_{i_z-i'_z}$ are
non-zero only for $|i_z-i'_z|\le \lambda$. The last term corresponds to the
last term in Eq.~(\ref{kinetic}). It is diagonal for the basis index in the $z$
direction due to its locality.

\section{The SO terms in various basis}
\label{Ham_so}

For the SO term, the matrix form can be obtained in a very similar way.
The SO part of the Hamiltonian reads
\begin{equation*}
-i\begin{pmatrix}
    \displaystyle
    \pdv{f(\vec{r})}{x}\pdv{}{y}-\pdv{f(\vec{r})}{y}\pdv{}{x} &
        \displaystyle \qty[\pdv{f(\vec{r})}{y}\pdv{}{z}-\pdv{f(\vec{r})}{z}\pdv{}{y}]-i\qty[\pdv{f(\vec{r})}{z}\pdv{}{x}-\pdv{f(\vec{r})}{x}\pdv{}{z}] \\
    \displaystyle
    \qty[\pdv{f(\vec{r})}{y}\pdv{}{z}-\pdv{f(\vec{r})}{z}\pdv{}{y}]+i\qty[\pdv{f(\vec{r})}{z}\pdv{}{x}-\pdv{f(\vec{r})}{x}\pdv{}{z}] &
    \displaystyle \pdv{f(\vec{r})}{y}\pdv{}{x}-\pdv{f(\vec{r})}{x}\pdv{}{y}
\end{pmatrix},
\end{equation*}
where $f(\vec{r})\equiv b_4 \rho(\vec{r})-b'_4 \rho_q(\vec{r})$ is a scalar 
function of position.

Next, we list the two typical terms in the SO term: $\displaystyle
\pdv{f(\vec{r})}{x}\pdv{}{y}$ and $\displaystyle \pdv{f(\vec{r})}{y}\pdv{}{z}$,
which contain operators with and without derivative in the $z$ direction:
\begin{eqnarray}
	\langle n_x n_y i_z | \pdv{f(\vec{r})}{x}\pdv{}{y} | n'_x n'_y i'_z \rangle 
 &=& \int dx \int dy \,\psi^*_{n_x}(x) \psi^*_{n_y}(y) \pdv{f(\vec{r})}{x} \psi_{n'_x}(x) \dv{\psi_{n'_y}(y)}{y} \delta_{i_z,i'_z}, \\
	\langle n_x n_y i_z | \pdv{f(\vec{r})}{y}\pdv{}{z} | n'_x n'_y i'_z \rangle 
 &=&  \int dx \int dy \,\psi^*_{n_x}(x) \psi^*_{n_y}(y) \pdv{f(\vec{r})}{y} \psi_{n'_x}(x) \psi_{n'_y}(y) m^{(1)}_{i'_z-i_z} (dz)^{-1}.
\end{eqnarray}

\bibliographystyle{apsrev4-2}
%

\end{document}